\documentclass[conference, a4paper]{IEEEtran}
\usepackage{simplemargins}
\usepackage{graphicx}
\usepackage{indentfirst}
\usepackage{epsfig}
\usepackage{amsfonts}
\usepackage{amsmath}
\usepackage{algorithmic}
\usepackage{algorithm}
\usepackage{comment}
\usepackage{xcolor}
\usepackage[utf8]{inputenc}
\usepackage[english]{babel}
\usepackage{amsthm}

\title{Optimal Fidelity-Aware Entanglement Distribution \\ in Linear Quantum Networks}
\author{Iordanis Koutsopoulos\\ Department of Informatics \\
Athens University of Economics and Business\\Athens, Greece}

\begin{document}

\maketitle
\newtheorem{property}{Property}
\newcommand{\be}{\begin{itemize}} \newcommand{\ee}{\end{itemize}}
\newcommand{\tb}{\textbf} \newcommand{\ttt}{\texttt}
\newcommand{\tit}{\textit} \newcommand{\uline}{\underline}
\newtheorem{proposition}{Proposition}
\newtheorem{theorem}{Theorem} \newtheorem{lemma}{Lemma}
\newtheorem{fact}{Fact}

\begin{abstract}
We study the problem of entanglement distribution in terms of maximizing a utility function that captures the total fidelity of end-to-end entanglements in a two-link linear quantum network with a source, a repeater, and a destination. The nodes have several quantum memories, and the problem is how to coordinate entanglement purification in each of the links, and entanglement swapping across links, so as to achieve the goal above. We show that entanglement swapping (i.e, deciding on the pair of qubits from each link to perform swapping on) is equivalent to finding a max-weight matching on a bipartite graph. Further, entanglement purification (i.e, deciding which pairs of qubits in a link will undergo purification) is equivalent to finding a max-weight matching on a non-bipartite graph. We propose two polynomial algorithms, the Purify-then-Swap (PtS) and the Swap-then-Purify (StP) ones, where the decisions about purification and swapping are taken with different order. Numerical results show that PtS performs better than StP, and also that the omission of purification in StP gives substantial benefits.
\end{abstract}

\section{Introduction}

The vision of the quantum internet \cite{survey0} is currently underway, and in the next years we will witness ground-breaking changes on a variety of areas applicable in tactical networks, such as secure information exchange using Quantum Key Distribution (QKD), improved sensing, and quantum processor networking for quantum computations. The quantum internet can be realized through terrestrial optical-fiber or wireless links, or satellite-to-ground-station links \cite{satellite}. The latter seem a good option, due to low attenuation of quantum signals through the atmosphere, and long
communication ranges. 

Regardless of the communication link, quantum communication is achieved through pairs of quantum bits (qubits), namely pairs of entangled photons, i.e, photons with a coupled state. Photon pairs are generated by a photon source and are distributed to two nodes at distinct locations through wireless or optical-fiber links. By measuring the information carried by one photon, it is possible to obtain information on the entangled one, based on the Einstein-Podolsky-Rosen (EPR) paradox of quantum mechanics. If two neighboring nodes, say Alice (A) and Bob (B) share a pair of entangled qubits, A can secretly send one information bit to B with the help of a quantum measurement, using the classical internet channel. This process is called \textit{teleportation}. Secret communication is achieved due to the laws of quantum mechanics for quantum entanglement that make it impossible to eavesdrop or measure the communicated bit. By repeating the process above, A and B can exchange a secret key with many information bits.

\begin{figure}[h]
\centering
\includegraphics[width=1.07 \linewidth]{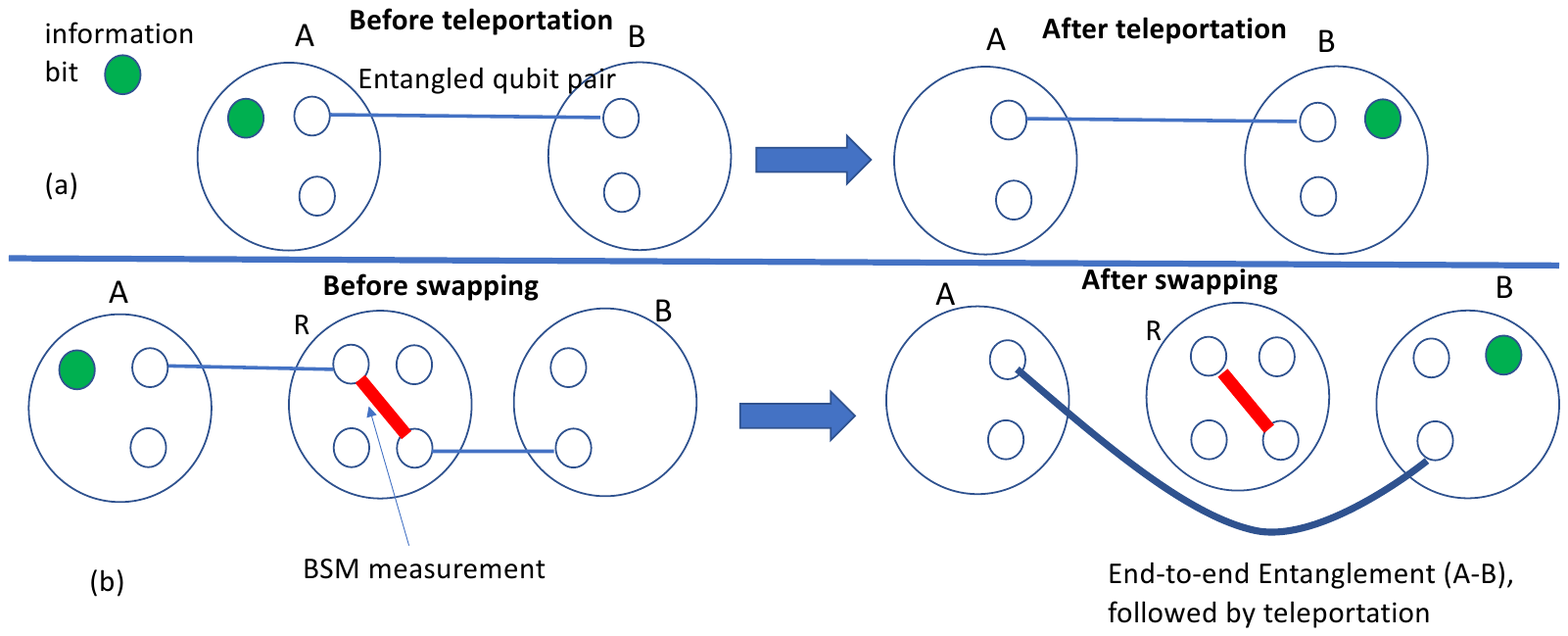} 
\vspace{-3cm}
\caption{(a) Quantum Teleportation, with which one bit of information is transmitted from A to B, if A,B share an entangled pair of qubits; (b) Entanglement swapping that leads to end-to-end entanglement between A and B, through BSM measurement at the qubits in the memories of repeater R. After end-to-end entanglement is established, A can send an information bit to B through teleportation.}
\label{fig:tele}
\end{figure}

Since transmitted signals over optical fibers or over the air attenuate with distance, the probability of successful entanglement between two nodes decreases at higher distance. Thus, repeater nodes are employed to maintain high probability of successful entanglement, while extending the distance range for the established entanglement. If nodes A and B need to share a pair of entangled qubits and they are not directly connected, a repeater node R that holds an entanglement to A through physical link $(A,R)$ and an entanglement to B through physical link $(R,B)$, can convert the two one-hop entanglements into one direct, longer-distance entanglement between A and B through \textit{entanglement swapping}, with a so-called Bell State Measurement (BSM) at node R. After swapping, the entangled pairs at each of the two links $(A,R)$ and $(R,B)$ do not exist, and an entangled pair between nodes A and B is created. Thus, a quantum repeater facilitates the transfer of quantum states between nodes attached to it through entanglement swapping. Swapping can be repeated in more intermediate repeaters to build entanglements over long distances. Teleportation and swapping are shown in Fig. \ref{fig:tele}.

Entangled qubits reside in quantum memories in repeaters. The quality of an entangled qubit pair is measured through its \textit{fidelity}, which captures at a given time the purity of quantum state, namely the similarity of the state of the entangled pair to its state when it was created. 
Stored qubits in quantum memories have limited lifetime since they are subject to \textit{decoherence}, which means that fidelity decreases with time, and after some time, they are rendered non-functional for quantum applications. 
Finally, two or more pairs of entangled qubits can be sacrificed to create one entangled pair of higher fidelity, and this process is known as entanglement \textit{purification}.

Entangled pairs of qubits are the most important resources in the quantum internet and are necessary for quantum protocols and procedures. A prime goal is to efficiently \textit{distribute a large number of high-fidelity entanglements} across the network, from source to destination. In a network in which nodes have several quantum memories, the main question is how to tune entanglement purification of stored pairs of qubits in quantum memories of the two links, and entanglement swapping across the two links so as to achieve the goal above.

We study the problem of efficient entanglement distribution in terms of maximizing a utility function that captures total fidelity of end-to-end entanglements in a two-link linear quantum network, where each node has multiple stored qubits in its quantum memories. At each time slot, a controller observes the system state, that is, the set of entangled pairs of qubits in each link, and their fidelity. We ask the question: given a continuous backlog of requests for end-to-end entanglement, how can we increase the derived utility through appropriate management of stored qubits in each link with entanglement swapping and purification? Our two-hop model may represent a terrestrial network with optical-fiber or wireless links, or one with satellite-to-ground links, and it is the simplest non-trivial model on which we show the joint impact of entanglement swapping and purification on entanglement distribution without involving entanglement routing. Also, this model stands for entanglement distribution over a general network topology and several source-destination pairs. If the route for each source-destination pair is given and routes do not intersect, then each route corresponds to the repeater chain model of this paper. The contributions of our work are as follows:
\begin{itemize}
\item We study the problem of increasing utility of end-to-end entanglement through purification and swapping operations in the stored qubits of quantum memories. We take into account essential assumptions such as decoherence of stored qubits with time, uncertain swapping and purification operations, and the impact of swapping and purification on the fidelity of entangled pairs.

\item We show that \textit{entanglement swapping} in the two-link network, i.e. deciding which pairs of qubits from each link to swap, is equivalent to finding a max-weight matching on a bipartite graph. Further, \textit{entanglement purification}, i.e, deciding which pairs of qubits of a link will undergo purification, is equivalent to finding a max-weight matching on a non-bipartite graph. 
 
\item We propose two polynomial algorithms, the Purify-then-Swap (PtS) and the Swap-then-Purify (StP) ones, where the decisions about entanglement purification and swapping are taken with different order. Numerical results show that PtS performs better than StP, and that the omission of purification in StP gives substantial benefits.

 \end{itemize}


Our work is the first one that studies the problem of entanglement distribution through nodes with multiple quantum memories, through the interplay of entanglement purification and swapping. The paper is organized as follows. In section \ref{section:related}, we provide a literature overview, while in section \ref{section:model} we present the model and problem statement, and in section \ref{section:solution}, we present two algorithms. Sections \ref{section:results} and \ref{section:conclusion} include numerical results and the conclusion respectively.

\section{Related work} \label{section:related}

\textbf{Quantum entanglement routing.} Quantum entanglement routing studies the distribution of entanglement over multiple hops. The most important metric is entanglement generation rate which measures the number of entangled pairs shared between source and destination per unit of time. In \cite{leandros-quantum1}, the authors 
propose entanglement 
distribution protocols under realistic assumptions on entanglement swapping, number and quality of quantum memories, link losses, and local link state information (e.g. in terms of entanglement success or failure). 

A survey of entanglement routing protocols is \cite{survey1}, with a view towards multiple commodities, where a commodity is a source-destination pair with an entanglement request. The work \cite{sigcomm20} is among the first that builds a quantum entanglement routing protocol by taking into account different topologies, link state exchanges, qubit capacity limitations, and commodities that compete for resources. The work \cite{elkouss} also studies a multi-commodity flow optimization problem to find the set of paths for each commodity such that the total end-to-end entanglement generation rate is maximized, subject to a minimum end-to-end fidelity for each commodity, reflected as a maximum path length constraint. In \cite{fid-guar}, the authors propose entanglement routing algorithms with purification decision to provide fidelity guarantees for many commodities.

In \cite{vardoyan1}, the authors use Markov Decision Processes to find the policy that minimizes delivery time of end-to-end entanglement in a repeater chain, where repeater memories have cut-off times due to decoherence. The optimal policy decides when to perform entanglement swapping in each repeater after observing the state, which is the age of each entangled qubit pair. The work \cite{cutoffs} also considers cut-off times. An entangled pair is kept for some time in the memory, and then it is discarded so that a new generated pair with higher fidelity is entered.
The authors optimize cut-offs to maximize end-to-end entanglement generation rate.
In \cite{wehner}, the authors design a quantum data plane protocol for end-to-end communication and provide mechanisms for coordinating entanglement swapping, handling decoherence of stored qubits, and ensuring high end-to-end fidelity through distillation. They also introduce the NetSquid simulator for quantum networks. 

The work \cite{qnum} defines three logarithmic utility functions that involve  entanglement generation rate and fidelity. Two of them respectively are interpreted as the utility from entanglement rate and distillation, the latter being a function of fidelity, and as the utility from the amount of bits in the secret key, which also involves the entanglement rate and entropy of Werner states. The third utility function is concave in entanglement rate and fidelity, thus mathematically tractable.

\textbf{Entanglement purification and routing}. The works \cite{tserkis} and \cite{victora} study the integration of purification protocols into entanglement distribution 
with realistic system constraints such as imperfect fidelities and limited storage time.
The works highlight the importance of tuning purification protocols for better end-to-end entanglement rate. In \cite{purification1}, the authors show that under perfect swapping and purification operations, the per-memory entanglement distribution rate increases as the number of memories increases even under memory decoherence; however this is not the case for imperfect operations. 

\textbf{Quantum switches}. The authors in \cite{towsley-stochastic} study a quantum switch that serves through its memories incoming bipartite end-to-end entanglement requests made by users users in a star topology. 
The authors use Continuous-Time Markov Chains to characterize the switch capacity, namely the maximum possible number of end-to-end entanglements served by the switch per time unit, as well as the probability distribution and expected number of qubits in the memories. Finally, the authors in \cite{promponas} propose a policy for allocating the limited quantum memories of a switch to link-level entanglements of clients under dynamic arrivals of entanglement requests. They characterize the switch capacity region, namely the set of request arrival vectors for which the queues of serving requests are stable and propose a memory allocation policy that is based on the max-weight algorithm \cite{leandros} and is throughput optimal.

\section{System Model and Problem Statement} \label{section:model}

\subsection{System model}

We consider a two-hop linear quantum network with a source node $s$, a repeater $r$, and a destination node $d$, and two physical links, $(s,r)$ and $(r,d)$. The network may represent a terrestrial system with optical-fiber or wireless physical links, or a hybrid terrestrial-satellite system, where nodes $s$, $d$ are ground stations, and the repeater $r$ is a satellite. Nodes $s$ and $d$ engage in applications such as QKD, thus entangled qubit pairs need to be generated and distributed end-to-end from $s$ to $d$ so that they are used in applications. Each of $s$ and $d$ has $M$ quantum memories, while the repeater has $2M$ quantum memories; $M$ of those are used to attempt link-level entanglements (LLEs) with the $M$ memories of $s$, while the other $M$ memories are used to attempt LLEs with the $M$ memories of $d$. 
Each memory can store one qubit. 

\subsubsection{Entanglement generation}

Time is slotted, and the slot duration $\delta$ is assumed to be larger than the time it takes for purification and swapping together to take place, and it can be of the order of tens of msecs. At each time slot, in link $(s,r)$, link-level entanglement (LLEs) are attempted as follows. Node $s$ has a photon generator that generates an entangled qubit (photon) pair. One photon is kept locally, and the other is sent to $r$ via an optical-fiber or free-space link. The probability of successful photon transmission through link $(s,r)$, $p_{sr}$, decays exponentially with distance due to photon loss. 

If the entangled qubit is received correctly, $r$ sends an acknowledgment message to $s$ through a classical communication channel, while one qubit of the entangled qubit pair is stored in a memory of $s$ and the other qubit in a memory of $r$. The entanglement generation process is repeated for $M$ times in parallel, and each successfully generated entangled qubit is stored at a memory $i$ of $s$, and the corresponding memory $i$ of $r$, for $i=1,\ldots,M$. Similarly, $M$ LLEs are attempted on link $(r,d)$ between memory $M+i$ of $r$ and memory $i$ of $d$, $i=1,\ldots,M$, with success probability is $p_{rd}$.

At each slot, the entanglement generation attempts in link $(s,r)$ are continual as long as there are unoccupied pairs of memories at node $s$ and $r$, and the indices of successful EPR pairs between $s$ and $r$ are known. The same holds for link $(r,d)$. If a pair of entangled qubits are shared by $s$ and $r$, then $s$ can send one bit of secret information to $r$ with 
teleportation.

\subsubsection{Fidelity and decoherence}

The quality of an LLE is measured by \textit{fidelity} $F$, which relates to the coherence of the two entangled qubits. When an LLE is generated, it has initial fidelity $F_0$. The quality of the LLE decays with time due to interactions of stored qubits 
with the environment, and this is known as decoherence. If the LLE has been generated at time slot $t_0$, its fidelity at time slot $t > t_0$ is \cite{survey1}, \cite{vardoyan1}:
\begin{equation}
F(t) = \frac{1}{4} + \left( F_0 - \frac{1}{4}\right) e^{-\delta(t-t_0)/\tau}
\label{eq:decay}
\end{equation}
where $\tau$ characterizes the exponential decay. If fidelity is smaller than a threshold, 
the qubit is 
useless for applications.

\subsubsection{Entanglement Swapping}

We assume bipartite entanglements. An end-to-end entanglement between nodes $s$ and $d$ can be created through entanglement swapping of one LLE of link $(s,r)$ and one LLE of link $(r,d)$. The swapping is realized with BSM of the two qubits at the corresponding memories of node $r$. The two LLEs disappear, and after swapping, there exists an entangled qubit pair between nodes $s$ and $d$. Entanglement swapping succeeds with probability $p_s$. If an end-to-end entanglement is successful, it comes with a fidelity that is a deterministic function of the fidelity values of the involved LLEs. We assume for simplicity that swapping operations are perfect, that is, Bell states preserve their form after swapping. If $F_1, F_2$ are the fidelity values of the two LLEs before swapping, the resulting end-to-end entanglement after swapping has fidelity:
\begin{equation}
F_s(F_1, F_2) =  F_1 F_2 + (1-F_1)(1-F_2)\,.
\end{equation}

\subsubsection{Entanglement Purification} 

The fidelity of entangled qubits (either LLEs or end-to-end) decreases with time. In order to increase fidelity, we use entanglement \textit{purification}. We assume $2$-to-$1$ purification, where two entangled qubit pairs in the same link 
are sacrificed to generate a single pair over that link, with higher fidelity than the sacrificed ones. In this case, the new purified qubit pair substitutes one of the two existing pairs in memories, while the other qubit pair is discarded and the memories are freed. We assume perfect purification operation which succeeds with probability $p_{pur} = F_1 F_2 + (1-F_1)(1-F_2)$ \cite{meter}. The fidelity of an entanglement resulting from purification of two entangled qubit pairs with fidelities $F_1$, $F_2$ is \cite{meter}
\begin{equation}
F_p(F_1, F_2) = \frac{F_1 F_2}{F_1 F_2 + (1-F_1)(1-F_2)}\,.
\end{equation}
After purification, the link nodes receive 
the fidelity value. We do not assume multiple levels of purification, namely once an LLE is formed from purification, it is not purified again.

\subsubsection{Sequence of events}

At the end of time slot $t-1$, entanglements are attempted between unoccupied memories of nodes $s$ and $r$, and between memories of $r$ and $d$ and succeed with probability $p_s$. At the beginning of time slot $t$, a controller at node $s$ observes the system state, namely the successful LLEs of links $(s,r)$ and $(r,d)$, and their fidelity values. Then, it decides on the order of purification and swapping, and on how they will be performed. Swapping and purification operations are realized during slot $t$, and $L$ attempts for purification may be made. After swapping and after purification, the resulting entanglement obtains the age of the oldest LLE involved in the operation. In the end of slot $t$, entanglements are again attempted between unoccupied memories of $s$, $r$ and those of $r$, $d$, and the process continues. At the end of each slot, the fidelity of each entangled qubit decays according to (\ref{eq:decay}). When a number of time slots elapses and fidelity drops below a threshold, the qubit pair is discarded, and the corresponding memories are emptied. 

\subsection{Problem statement}

Since nodes have $M$ quantum memories, we can create at most $M$ 
entanglements from $s$ to $d$. For quantum applications, it is important to have a large number of 
entangled qubits so that we securely transmit more bits. On the other hand, the fidelity of entangled qubits needs to be high, so that they can better support applications, and for longer time before they decohere. Let $\mathbf{x}$ be the $M \times 1$ binary vector of end-to-end entanglements and $\mathbf{F}$ the vector of fidelity values, where $x_i = 1$ means that there exists an end-to-end entanglement that originates from the $i$-th memory pair of $(s,r)$, and $0$ if not, and $F_i$ is the fidelity of this entanglement. The importance of the number and fidelity values of entanglements is captured by a utility function
\begin{equation}
u(\mathbf{x},\mathbf{F}) = \displaystyle \log(\sum_{i=1}^M x_i g(F_i))\, 
\label{eq:utility}
\end{equation}
\begin{figure}[h]
\centering
\includegraphics[width=1.1 \linewidth]{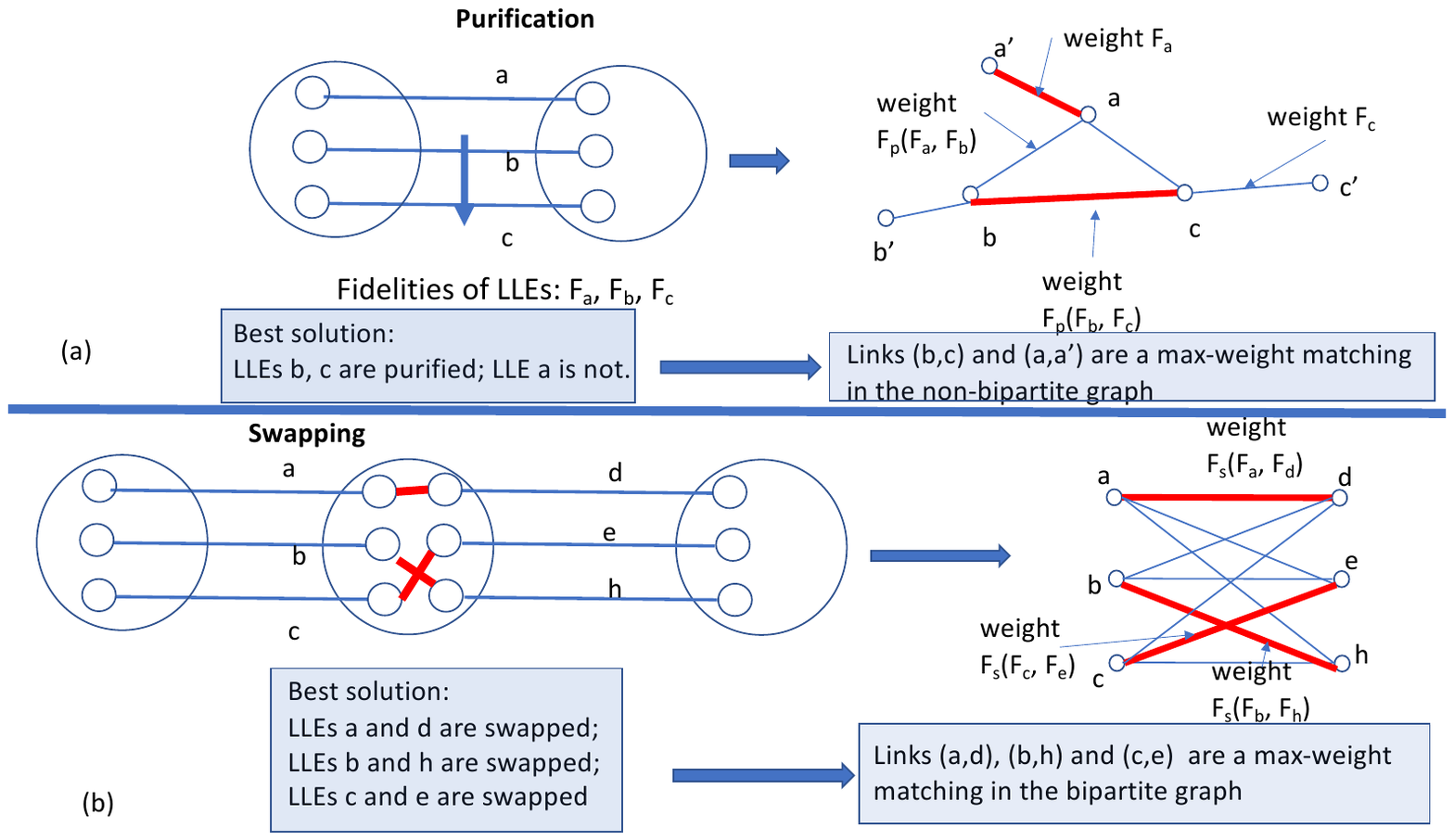} 
\vspace{-1.5cm}
\caption{Upper part (a): Equivalence of the optimal purification problem to max-weight matching in a non-bipartite graph. We show a hypothetical best solution: LLEs b and c are purified, while LLE a is not. This corresponds to a max-weight matching in the non-bipartite graph on the right. 
Bottom part (b): Equivalence of the optimal swapping problem to max-weight matching in a bipartite graph. We show a hypothetical best solution: LLEs a,b are swapped, and so are LLEs b,h, and LLEs c,e. This corresponds to the max-weight matching in the bipartite graph on the right.}
\label{fig:Matching}
\end{figure}
where function $g(\cdot)$ depends on the application \cite{qnum} and denotes utility out of fidelity. On the one hand, it is important to perform the swapping of qubit pairs from links $(s,r)$ and $(r,d)$ so as to have high total fidelity. 
On the other hand, purification leads to a smaller number of entangled pairs, albeit of higher fidelity. In order to achieve good utility, we need to balance the trade-off above by managing the swapping and purification operations. It is not clear whether we should strive for a small number of entangled qubit pairs of large fidelity or for a higher number of qubit pairs of lower fidelity. The order of swapping and purification plays an important role.

\section{Entanglement Swapping and Purification} \label{section:solution}

We propose two algorithms for high utility of end-to-end entanglements. The algorithms differ in the order they perform swapping and purification. First, we show how to do each of these operations efficiently.

\subsection{Optimal Entanglement Swapping} \label{subsection:sw}

First, consider entanglement swapping between LLEs of link $(s,r)$ and LLEs of link $(r,d)$ in order to form end-to-end entanglements with high utility. At each time slot, the controller sees the 
successful entangled qubit pairs and their fidelity values in each of the two links. We define a bipartite graph $\mathcal{G} = (\mathcal{U} \cup \mathcal{V}, \mathcal{E})$. Each node $i \in \mathcal{U}$ stands for an LLE in link $(s,r)$, namely a pair of memories $(i,i)$ with a successful entangled qubit pair (one memory in node $s$ and another in $r$). Each node $j \in \mathcal{V}$ stands for an LLE in link $(r,d)$, namely a pair of memories $(M+i,i)$ with a successful entangled qubit pair (one memory in node $r$ and the another in $d$). Let $F_i, F_j$ be the fidelity values of those LLEs. We define the weight of link $(i,j)$ as $w_{ij} = g(F_s(F_i, F_j))$, which is the utility of
fidelity $F_s(F_i, F_j)$ after swapping between LLEs $i$ and $j$. 

The problem of finding the LLE swapping operation that maximizes total utility of end-to-end entanglements is equivalent to finding a \textit{max-weight matching} $\mathcal{M}^*$ in bipartite graph $\mathcal{G}$. The edges of the matching denote the LLEs of links $(s,r)$ and $r,d)$ that will be swapped (Fig. \ref{fig:Matching}(b)). The max-weight matching can be found with the Hungarian algorithm \cite{hungarian}. 

\subsection{Optimal Entanglement Purification}

Next, we consider $2$-to $1$ purification of LLEs in link $(s,r)$ (and similarly for $(r,d)$), namely finding the pairs of LLEs to be merged so that the resulting LLEs have high utility. Focus on link $(s,r)$. We define the non-bipartite graph $\mathcal{G}_{sr} = (\mathcal{V}_{sr},\mathcal{E}_{sr})$. The node set $\mathcal{V}_{sr}$ consists of one node $i$ for each successful LLE of $(s,r)$, and one replica node $i'$ of each node $i$. The set of edges $\mathcal{E}_{sr}$ is formed as follows. We connect nodes $i$ and $j$, $i \neq j$, that stand for successful LLEs. We assign to link $(i,j)$ weight $w_{ij} = F_p(F_i, F_j)$, which is the utility after purification of LLEs corresponding to $i$ and $j$, with initial fidelity $F_i$ and $F_j$. We also connect each node $i$ only with its own replica node $i'$, and we assign to link $(i,i')$  weight $w_{ii'} = F_i$. This link denotes that the LLE corresponding to $i$ will not be purified, thus its fidelity remains the same. Thus, links $(i,j)$ with $i \neq j$ stand for the tentative pairing of LLEs $i$ and $j$ for purification, while links $(i,i')$ stand for tentative non-pairing of LLE $i$ with another LLE. 

The problem of purification on link $(s,r)$ to maximize utility is equivalent to finding a \textit{max-weight matching} $\mathcal{M}^*_{sr}$ on $\mathcal{G}_{sr}$. The edges in $\mathcal{M}^*_{sr}$ dictate the way to perform the purification: edges $(i,j)$, $i \neq j$, in $\mathcal{M}^*_{sr}$ mean that LLE $i$ is merged with LLE $j$ and purified; edges $(i,i')$ in $\mathcal{M}^*_{sr}$ mean that LLE $i$ is not merged with another LLE and maintains its fidelity (Fig. \ref{fig:Matching}(a)). The max-weight matching can be found in polynomial time with the Edmonds blossom algorithm \cite{edmonds}. 

\subsection{Purify-then-Swap algorithm}

The Purify-then-Swap (PtS) algorithm performs first purification and then swapping. For the purification of LLEs on link $(s,r)$, we find a max-weight matching $\mathcal{M}^*_{sr}$ in the non-bipartite graph $\mathcal{G}_{sr}$ defined above. Similarly, we purify LLEs on link $(r,d)$ by finding a max-weight matching $\mathcal{M}^*_{rd}$ in the corresponding graph $\mathcal{G}_{rd}$. The set of edges in each matching denotes for each link the resulting purified LLEs after merging, and the LLEs that will not undergo purification. 

\begin{figure}[h]
\centering
\includegraphics[width=1.18 \linewidth]{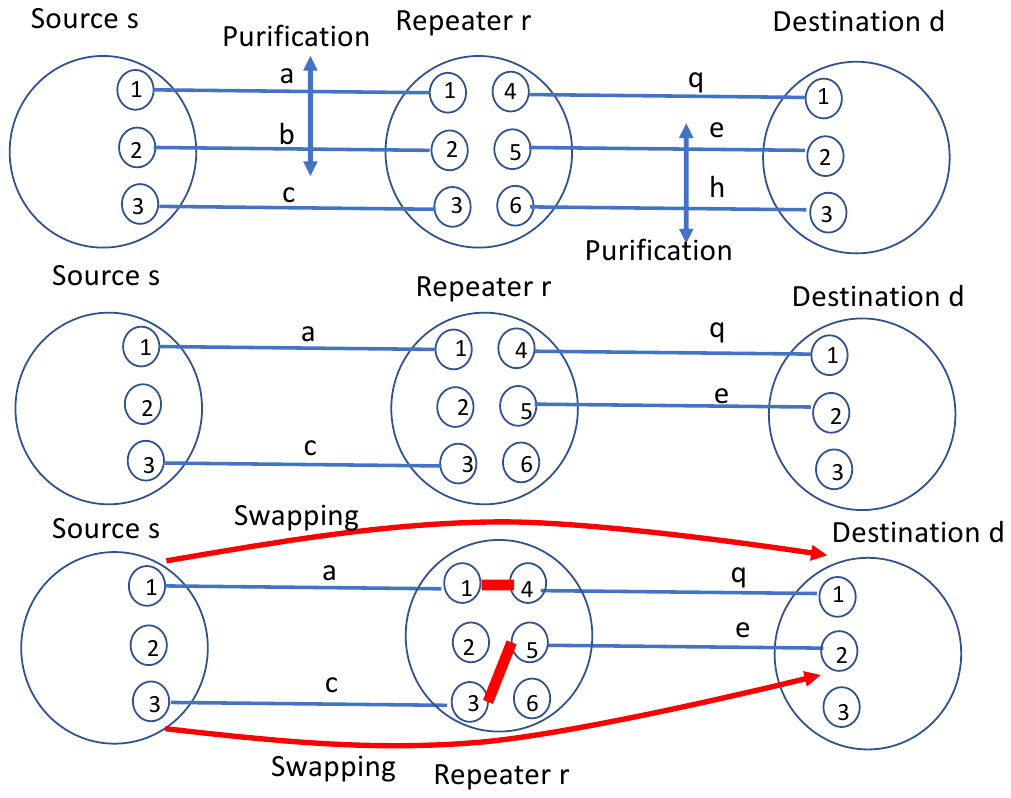} 
\vspace{-2cm}
\caption{Example of the Purify-then-Swap algorithm. Purification: In link $(s,r)$, LLEs a and b are purified, while c is not. The new purified LLE is stored in memory 1 of s and 1 of r. In link $(r,d)$, LLEs e and h are purified, while q is not. The new purified LLE is stored in memory 4 or r and 2 of d. Swapping:LLEs a and q are swapped, and also LLEs c and e are swapped. The outcome is two end-to-end entanglements.}
\label{fig:Purify-Swap}
\end{figure}

Next, we define bipartite graph $\mathcal{G}_s = \mathcal{U}_s \cup \mathcal{V}_s, \mathcal{E}_s$ as follows, for the swapping. For each link in $\mathcal{M}^*_{sr}$, we create a node $k \in \mathcal{U}_s$, and for each link in $\mathcal{M}^*_{rd}$, we create a node $\ell \in \mathcal{V}_s$. We assign a weight to link $(k,\ell)$ as follows. If node $k$ corresponds to a matched edge of the form $(i,j)$, $i \neq j$ for link $(s,r)$, and node $\ell$ corresponds to a matched link of the form $(i_0, j_0)$, $i_0 \neq j_0$ for link $(r,d)$, the weight is
$w_{k \ell} = g(F_s(F_p(F_i, F_j), F_p(F_{i_0}, F_{j_0})))$, which corresponds to swapping of LLEs $k$ and $\ell$ that have emerged through purification. If node $k$ corresponds to a matched edge of the form $(i,i')$, for link $(s,r)$, and node $\ell$ corresponds to a matched edge of the form $(i_0, i'_0)$ for link $(r,d)$, the weight is $w_{k \ell} = g(F_s(F_i, F_{i_0}))$. The weights for the other two cases of links are assigned similarly. The LLEs that will be swapped result from a max-weight matching on $\mathcal{G}_s$. An example of the PtS algorithm is shown in Fig. \ref{fig:Purify-Swap}.

\subsection{Swap-then-Purify algorithm}

In the Swap-then-Purify (StP) algorithm, swapping is performed first by finding a max-weight matching $\mathcal{M}^*$ on bipartite graph $\mathcal{G}$ described in subsection \ref{subsection:sw}. The edges in the matching denote the pairs of LLEs, one from link $(s,r)$ and one from link $(r,d)$ that will be swapped. Let $\mathcal{N}$ be the set of end-to-end entanglements created after swapping. We index these entanglements as $n=1,\ldots, |\mathcal{N}|$ and denote by $F_n$ the fidelity of the $n$-th end-to-end entanglement after swapping.

Next, we form the non-bipartite graph $\mathcal{G}_p$ with one node $n$ for each end-to-end entanglement, and one replica node $n'$ for each $n$. We connect node $n$ with each other node $m \neq n$, and we connect $n$ with $n'$. To each edge $(n,m)$, $m \neq n$, we assign weight $w_{mn} = g(F_p(F_n, F_m))$, which is the fidelity after purification of the $n$-th and $m$-th end-to-end entanglements. To each edge $(n,n')$, we assign weight $w_{nn'} = g(F_n)$, as this denotes entanglements that are not purified. The max-weight matching $\mathcal{M}^*_p$ on $\mathcal{G}_p$ denotes the way that pairs of end-to-end entanglements will be purified or not.

\section{Numerical results} \label{section:results}

We consider a two-hop linear network $s \rightarrow r \rightarrow d$, where $s$, $d$ have $M=3$ memories each, and $r$ has $6$ memories. We define two utility functions for the utility of the $n \leq 3$ end-to-end entanglements \cite{qnum}: 
\begin{itemize}
\item $u_A(\mathbf{F}) = \log (\sum_{i=1}^n F_i)$. 
The utility is the logarithm of the total fidelity of entanglements between $s$ and $d$. 
\item $u_B(\mathbf{F}) = \log(\sum_{i=1}^n D(F_i))$, where  $D(\cdot)$ measures the rate at which perfect Bell states are distilled from bipartite states and is given by
\begin{equation}
D(x) = 1 + x \log_2 x + (1-x) \log_2\frac{1-x}{3}\,.
\end{equation}
\end{itemize}

As discussed in \cite{qnum}, $u_A(\cdot)$ and $u_B(\cdot)$ are appropriate for applications with emphasis on many entanglements and high fidelity respectively.
We compare the following algorithms: (\textit{i}
Purify-then-Swap (PtS); (\textit{ii}) Swap-then-Purify (StP); (\textit{iii}) Swap-only, with no purification after swapping.

We carried 500 experiments; in each experiment, we created 6 LLES: 3 on link $(s,r)$ and 3 on link $(r,d)$, with fidelity values uniformly distributed in $[0.8, 1]$. Figures \ref{fig:Ut1} and  \ref{fig:Ut2} show the comparative results of the 3 algorithms for utility functions $u_A(\cdot)$ and $u_B(\cdot)$ respectively. Interestingly, the Swap-only approach achieves better perfomance than PtS and StP. Also, for $u_A(\cdot)$, PtS achieves $4-5 \%$ better performance than StP, suggesting that it is better to do purification first. For $u_B(\cdot)$, PtS is $20\%$ better than StP, and Swap-only is again best of all. 
The additive nature of utility functions favors more entanglements, and thus it favors PtS over StP. The superiority of Swap-only is again due to the additive utility functions.  

\section{Conclusion} \label{section:conclusion}

We studied the interplay of swapping and purification towards achieving high utility of end-to-end entanglements in a two-hop linear network. First, we showed that the optimal swapping and purification problems can be viewed as max-weight matching problems in bipartite and non-bipartite graphs respectively. Then, we proposed two algorithms that execute these operations in different order. This simple setting offers some first insights which are seemingly due to the nature of the utility function that is additive in the fidelity values of entanglements. These insights are the superiority of Swap-only poilcy, and the better performance of PtS over StP. 

The methods can be extended to larger repeater chains and can be viewed jointly with routing i.e, determination of the repeater path that forms the end-to-end entanglement. Further, new utility functions should quantify the priority between the number of entanglements and their fidelity, depending on the application. For example, if the application requires handling large amounts of data or can tolerate higher error rates with good error correction methods, more entanglements of lower fidelity might make more sense, while in other cases priority could be placed on higher fidelity.
Imperfections in purification and swapping can also be modeled, as well as generalizations of purification so as to substitute $m>2$ LLEs with $n<m$ entanglements of higher fidelity, where $n$ is to be decided so that we free different number of memories.
Another question whether deciding the purification and swapping jointly rather than sequentially is better. Finally, in the general case, with dynamic requests for end-to-end entanglements between different source-destination pairs, the problem is to manage repeater memories, purify LLEs of links, find the route and form end-to-end entanglements so as to maximize the utility, while keeping the queue of requests stable.

\end{document}